\documentclass[preprint2]{aastex}
\newcommand{\hd}{HD~209458}
\newcommand{\hdb}{HD~209458b}
\newcommand{\au}{\textsc{au}}
\newcommand{\Rjup}{\ensuremath{R_{\mathrm{Jup}}}}
\newcommand{\Mjup}{\ensuremath{M_{\mathrm{Jup}}}}
\newcommand{\Ptrans}{\ensuremath{\mathcal{P}_\mathrm{transit}}}
\newcommand{\periods}{\mbox{\ensuremath{10\mathrm{d}<P<200\mathrm{d}}}}
\newcommand{\nsig}[1]{\mbox{{#1}-$\sigma$}}

\shorttitle{Monte Carlo simulations of \url{transitsearch.org}}
\shortauthors{Seagroves et al.}

\slugcomment{submitted to PASP 16 September 2003, accepted 2 October 2003}

\begin{document}

\title{Detection of Intermediate-Period Transiting Planets with a Network of Small Telescopes: \lowercase{\url{transitsearch.org}}}

\author{Scott Seagroves, Justin Harker, Gregory Laughlin, Justin Lacy}
\affil{UCO/Lick Observatory, Dept.\ of Astronomy \& Astrophysics, 
UC Santa Cruz, Santa Cruz CA 95064}
\email{scott@ucolick.org, jharker@ucolick.org, laugh@ucolick.org, lacy@transitsearch.org}

\and

\author{Tim Castellano} 
\affil{Astrophysics Branch, MS 245-6, NASA Ames Research Center,
  Moffett Field CA 94035}
\email{tcastellano@mail.arc.nasa.gov}

\begin{abstract}
We describe a project (\url{transitsearch.org}) currently attempting to
discover transiting intermediate-period planets orbiting bright parent
stars, and we simulate that project's performance.  The discovery of such
a transit would be an important astronomical advance, bridging the
critical gap in understanding between \hdb\ and Jupiter.  However, the
task is made difficult by intrinsically low transit probabilities and
small transit duty cycles.  This project's efficient and economical
strategy is to photometrically monitor stars that are known (from
radial velocity surveys) to bear planets, using a network of
widely-spaced observers with small telescopes.  These observers, each
individually capable of precision (1\%) differential photometry,
monitor candidates during the time windows in which the radial
velocity solution predicts a transit if the orbital inclination is
close to $90\degr$.  We use Monte Carlo techniques to simulate the
performance of this network, performing simulations with different
configurations of observers in order to optimize coordination of an
actual campaign.  Our results indicate that \url{transitsearch.org} can
reliably rule out or detect planetary transits within the current
catalog of known planet-bearing stars. A distributed network of
skilled amateur astronomers and small college observatories is a
cost-effective method for discovering the small number of transiting
planets with periods in the range \periods\ that orbit bright
\mbox{($V<11$)} stars.
\end{abstract} 


\section{Introduction}\label{intro-sec}

Over the past seven years, Doppler radial velocity (RV) measurements
have led to the discovery of over one hundred planets within a sample
of several thousand bright, nearby Sun-like stars. As the catalog of
worlds continues to grow, our view of extrasolar planets is shifting
from an anecdotal collection of individual systems, e.g.\ 51 Pegasi,
$\upsilon$ Andromedae, or 47 Ursae Majoris, to a more complete
statistical census, in which categories and populations of planets can
be clearly delineated \citep{mcm2000}.\footnote{An up-to-date version
of the planetary census can be found at
\url{http://cfa-www.harvard.edu/planets/}} Yet the planetary systems
from which we can learn the most --- those that transit --- remain
anecdotal at best.

For each system, there is a chance that the planet will periodically
transit the surface of the star as seen from Earth. An eclipsing
Jupiter-mass planet on a 3-day orbit produces a periodic
\mbox{$\sim$1.5\%} dimming of the parent star that lasts for about 3
hours. At present (August 2003) only a single transiting planet ---
\hdb, $P=3.525\mathrm{d}$ --- has been studied in detail
\citep{cblm2000,hmbv2000}, while a second object (OGLE TM-56-b,
\citet{ktjs2003}) has been recently announced, but not studied as
extensively. \hdb\ has provided a scientific bonanza, including direct
and accurate measurements of the planet's radius ($1.35 \pm 0.06
\Rjup$; \citet{bcgnb2001}), mass ($0.69 \pm 0.05 \Mjup$;
\citet{m+2000}), density, and even sodium in its atmosphere and
hydrogen in its exosphere \citep{cbng2002,vm+2003}.

The excitement generated by \hdb\ has led to a major push by the
community to find additional transiting planets. A website maintained
by Keith
Horne\footnote{\url{http://star-www.st-and.ac.uk/\~{}kdh1/transits/table.html}}
lists, along with the project described in this paper, an additional
twenty-four ground-based collaborations that are engaged in various
efforts to discover planetary transits. In total, these surveys yield
a reported capacity for discovering 148 planets per month. Despite
this activity, however, an important corner of parameter space
receives extremely little coverage; there is currently no other
organized effort to detect intermediate-period planets which transit
bright ($V<11$ parent stars). We describe a strategy for detecting
such transits, which we have adopted for the \url{transitsearch.org}
collaboration, and we show Monte Carlo simulations that demonstrate
the project's feasibility.

Our basic approach is to harness a network of small independent
telescopes to obtain multiple differential-photometric time-series of
\textit{known} planet-bearing stars during the well-defined time windows
in which transits are predicted to occur. If several independent
observers simultaneously measure a characteristic diminution or
brightening at the predicted times of ingress or egress, then there is
strong evidence that the star is exhibiting a transit, and follow-up
confirmation can then be obtained at the time of the next predicted
transit. The observational campaign is coordinated through a website:
\url{www.transitsearch.org}.

\section{Scientific Motivation}\label{motive-sec}

Any transits which our network uncovers will occur for planets which
occult bright ($V<11$) stars. This is an advantage. Such stars are
precisely those for which the RV method can provide accurate orbital
parameters and accurate values for \mbox{$M \sin(i)$}, both of which
are required to usefully characterize the planetary
properties. Furthermore, a bright parent star facilitates accurate
photometry. The exquisite precision \mbox{($1.1\times10^{-4}$)} HST light
curves which have been produced for \hdb\ \citep{bcgnb2001} depend on
the \mbox{$V=7.64$} magnitude of the parent star. \citet{bcgnb2001}
report that in order to obtain optimal photon noise-limited precision
with HST for \hdb, photometric measurements of 80s duration (60s
integration plus 20s CCD readout) were required. The critical ingress
and egress periods were thus time-resolved into approximately 20
samples each. A \mbox{$V\sim9$} star, which produces \mbox{$\sim$6}
times fewer photons, would require 6-minute cadencing to obtain the
same photometric precision, and the periods of ingress and egress
would be resolved into only a few time intervals. For considerably
dimmer stars (\mbox{$V\sim14$,} say) photometric precision will
necessarily be compromised.

The \url{transitsearch.org} collaboration is geared to survey planets
with \periods. This sensitivity to longer-period transits occurs
because we can narrow our observations to specific predicted time
windows.  The detection strategy thus involves no data folding, and
does not demand stable photometry over multiple nights or seasons.

Why would an intermediate-period transiting planet be of interest?
Although the measured \mbox{($1.35\Rjup$)} radius of \hdb\ is
broadly consistent with its being a gas-giant composed primarily of
hydrogen \citep{gbhls1996,b+2000}, recent work by
\citet{gs2002,bll2003,bcbah2003} all suggest that our understanding of
irradiated giant planets is incomplete. These three
studies agree that standard evolutionary models can recover the
observed radius of \hdb\ only if the deep atmosphere is
unrealistically hot. The recent studies incorporate
realistic atmospheric temperature profiles; the no-core models of
\hdb\ have a radius of \mbox{$\sim$$1.1 \Rjup$}, which is much too
small.

Three resolutions to this problem have been suggested. \citet{bll2003}
show that \hdb\ might be receiving interior tidal heating through
ongoing orbital circularization resulting from perturbations due to a
second planetary companion, whereas \citet{gs2002} propose
that strong insolation-driven weather patterns on the planet are
leading to conversion of kinetic wind energy into thermal energy at
pressures of tens of bars. \citet{bsh2003} argue that the size
discrepancy stems largely from improper interpretation of the transit
radius, and that the measured radius of \hdb\ in fact lies much higher
up in the planetary atmosphere than is generally assumed.

In any event, an accurate size and mass determination for an
intermediate-period planet will be of great help in resolving the
observed size discrepancy for \hdb.  A planet with intermediate period
cannot not have significant internal tidal dissipation, but would
still be receiving a modest amount of kinetic heating from the
mechanism suggested by \citet{gs2002}. Furthermore, the
\citet{bsh2003} theory for \hdb\ is readily extended to predict
effective transit radii at different planetary masses and
temperatures; the discovery of an intermediate period transiting
planet would provide a useful test of such predictions.

Intermediate-period planets are also interesting because they can
harbor dynamically stable large satellites. Tidal interactions likely
removed any satellites larger than \mbox{$R=70\mathrm{km}$} orbiting
\hdb.  Mars-mass moons, however, can last for 5~Gyr in the Hill Sphere
of a \mbox{$1 \Mjup$} planet orbiting a \mbox{$1 M_{\sun}$} star in a
27 day \mbox{(0.18 \au)} orbit, whereas in a 54 day \mbox{(0.28 \au)}
orbit, Earth-mass moons are dynamically stable \citep{bo2002}.
\citet{bcgnb2001} report that with HST, detections of satellites as
small as \mbox{$1 R_{\oplus}$} are feasible. Therefore, an
intermediate-period planet found by our survey could be followed up to
search for large moons, and, additionally, planetary rings.  Prior to
space-based missions such as KEPLER \citep{boru2003}, the detection of
a large moon orbiting an intermediate-period transiting planet is the
best prospect for finding a habitable world.

\section{How Many Planets Transit Bright Stars?}\label{how-many-sec}

Regardless of scientific benefit, our survey can be successful only if
there are additional transiting planets to be found orbiting bright
stars, and only if the telescope network is sensitive and responsive
enough to definitively confirm or rule out the occurrence of transits
for individual stars.

The \textit{a priori} probability that a planet transits its parent
star as seen from the line of sight to Earth is given by
\begin{equation}\label{Ptrans-eq}
\Ptrans = 0.0045 \,
\frac{1 \au}{a} \,  \frac{R_{\star}-R_{\mathrm{p}}}{R_{\sun}} \, 
\frac{1+e \cos(\pi/2-\varpi)}{1-e^{2}}  \, ,
\end{equation}
where $a$ is the semi-major axis of the orbit, $R_{\star}$ is the
radius of the star, $R_{\mathrm{p}}$ is the radius of the planet, $e$
is the orbital eccentricity, and $\varpi$ is the argument of
periastron referenced to the plane of the sky. Using the parameters of
the current radial velocity planet catalog\footnote{see, e.g.,
\url{http://www.transitsearch.org/stardatabase/index.htm}}, we find
that among the 17 Doppler wobble planets with periods \mbox{$P<10
\mathrm{d}$}, there are \mbox{$\langle n \rangle=1.75$} expected
transits, and indeed, within this group, a transiting case (\hdb) is
known. Sixteen of the planets with \mbox{$P<10 \mathrm{d}$} have
reported non-detections (although in some cases unpublished and
unverified).  These non-detections include HD~68988b, HD~168743b, and
HD~217107b, which can be ruled out on the basis of observations made
with the \url{transitsearch.org} network.  Only one
\mbox{$P<10\mathrm{d}$} planet, HD~162020b \mbox{($P=8.428
\mathrm{d}$)}, has, to our knowledge, not yet been checked for
transits.

Among the aggregate of 27 planets having periods in the range
\periods, the expected number of transiting planets is \mbox{$\langle
n \rangle=0.72$}.  Almost none of the parent stars in this group,
however, have yet been monitored for transits, due to low individual
transit probabilities and increasingly uncertain transit
ephemerides. The main sequence stars harboring intermediate-period
planets therefore represent the primary targets for our network. We
also note that among the 67 known planets with
\mbox{$P>200\mathrm{d}$}, one expects \mbox{$\langle n \rangle\approx
0.6$} additional transiting cases.  However, as the planetary period
becomes longer, follow-up becomes increasingly difficult due to
uncertainties in the transit times and long intervals between
occultations.

The transit probability for any given planet is not a strictly
declining function of semi-major axis. For example, the highest
\textit{a priori} transit probability for any known planet belongs not
to one of the short-period hot Jupiters (which tend to average
\mbox{$\Ptrans \sim 12\%$}), but rather to the
\mbox{$P=550\mathrm{d}$} planet orbiting $\iota$~Draconis.  In this
system, a large planetary semi-major axis (\mbox{1.34 \au}) is more
than offset by the \mbox{$12.8 R_{\sun}$} stellar radius
\citep{al1999}, and favorable orbital geometry (\mbox{$e$=0.7},
\mbox{$\varpi=94 \degr$}; \citet{fmqfmb2002}) which lead to
\mbox{$\Ptrans = 15.4\%$}, with the next predicted transit occuring on
April 4, 2004.  An extreme case such as this leads to a transit depth
which is hard to detect from the ground (and impossible for our
network) but there are other intermediate-period planets which have
surprisingly high transit probabilities (e.g. HD~38529b:
\mbox{$P=14.5\mathrm{d}$}, \mbox{$\Ptrans=13.7\%$}, or HD~74156b:
\mbox{$P=51.6\mathrm{d}$}, \mbox{$\Ptrans=4.3\%$}).

In addition to the current census, more planets with periods suitable
for \url{transitsearch.org} will emerge if RV surveys expand their samples.
Currently, within the \periods\ range, there are five known planets
orbiting stars with \mbox{$V<6$}, seven orbiting stars with
\mbox{$6<V<7$}, six orbiting stars with \mbox{$7<V<8$}, and two
planets each in the \mbox{$8<V<9$}, and \mbox{$9<V<10$}
ranges. If we assume that every available chromospherically
quiet main sequence dwarf with \mbox{$V<6$} has been adequately
surveyed for \mbox{$P<200\mathrm{d}$} planets, and that each magnitude
bin of unit width contains 1.8$\times$ as many stars as available for
bin \mbox{$(V-1)$} \citep{allen2000}, then we expect that roughly
\mbox{$9+16+29+52+94=200$} detectable planets with \mbox{$P<200
\mathrm{d}$} exist in orbit around stars with \mbox{$V<11$},
indicating that close to 180 additional planets in this category can
be detected using current RV techniques for bright stars.
Statistically, this implies that 6 intermediate-period transiting
planets orbit bright nearby stars.\footnote{ A similar calculation
shows that 6 additional {\it short-period} \mbox{$(P<10\mathrm{d})$}
transiting planets are likely to be orbiting bright stars.} The goal of
the \url{transitsearch.org} network is to find one of these transits.

\section{Transit Detection With Small Telescopes}\label{small-telescopes-sec}

In the past several years, a number of amateur astronomers have
detected the \hdb\ transits and have shown that the \mbox{$\sim 1\%$}
diminution produced by a transiting Jovian planet is readily
observable via differential photometry obtained with small (8-10 inch
aperture) telescopes fitted with commercial-grade CCD detectors.  A
report of one of these observations (Arto Oksanen, Sky \& Telescope,
January 2001) raised a provocative question: Is it a realistic
possibility for a network of small-college observatories and highly
experienced amateurs to discover a new transiting system? If so, many
small telescopes can be organized to maintain a time-intensive
volunteer-based transit survey of known planet-bearing stars during
predicted transit epochs.

In order to investigate the viability of detecting transits using
low-cost equipment and software, we designed and documented an
end-to-end procedure which allowed us to observe an \hdb\ transit.
Our demonstration observatory consists of a Meade LX-200 8-inch f/10
telescope fitted with a Santa Barbara Instruments Group ST7E 765x510
pixel CCD. Pointing/imaging/guiding, standard image reductions, and
aperture photometry are accomplished with a laptop computer running
The Sky, CCDSoft, and MIRA AP 6.0 software, respectively.  These tools
are all well-documented, reliable, relatively inexpensive, and
familiar to amateur astronomers.  

\begin{figure}[t]
\plotone{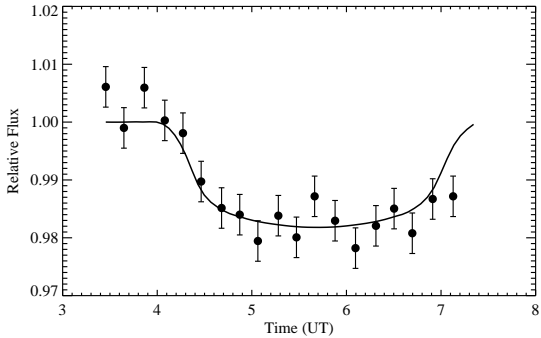}
\caption{
Detection of the planet transiting \hd, using the
portable observatory described in the text. The data were taken from
Fremont, CA on the night of 19/20 October 2001.  The solid curve is
the model of \citet{bcgnb2001}. 
\label{transit-fig} 
}
\end{figure}

With a focal reducer, the CCD image of a target region covers
\mbox{$36\arcmin \times 24\arcmin$}, which is generally large enough
to admit several \mbox{$V=9-11$} comparison stars for differential
photometry.  In the case of \hd, a field star HIP~108793
\mbox{($V=8.33$)} is situated $12\arcmin$ away. We acquire the target
field prior to the predicted start of ingress, and obtain successive
2s CCD exposures at a cadence of 35s per frame. The short 2s exposures
are used to avoid pixel saturation by \hd. The small overall duty
cycle is caused by the need to acquire auto-guiding images and to read
out the CCD.  Sequences of 20 exposures are averaged together to
produce composite measurements of the brightness of the stars in the
field within 12 minute intervals. Using standard aperture photometry
techniques, photoelectron counts from the target star are compared to
counts from the comparison star(s) in the field. A transit manifests
itself by the characteristic changes in the brightness of the target
star during the predicted times of ingress and/or egress. This
phenomenon is shown in Figure~\ref{transit-fig}, for \hdb\ during the
transit of 10/19/01. Photometric errors for each 12 minute bin are of
order 0.003 magnitudes, and are dominated by atmospheric
scintillation.  We note that simple improvements, such as the use of a
broad band, neutral density, or spot filter \citep{cast2000} to
increase open-shutter time or reduce the difference in brightness
between the target and a typical comparison star, could considerably
improve precision.

For our purposes here, this observation tells us two things. First,
transiting planets are readily detected with standard amateur-oriented
equipment. Second, we can assume that many observers worldwide will
have similar observational configurations, and will be capable of
obtaining differential photometry of comparable precision.

\section{Monte Carlo Simulation}

In order to evaluate the viability of a collaboration-based transit
survey of known planet-bearing stars, we have performed a Monte Carlo
study which models realistic incarnations of the
\url{transitsearch.org} network.  This simulation is written in
IDL\footnote{\url{http://www.rsinc.com}} and makes heavy use of the IDL
Astronomy User's Library\footnote{\url{http://idlastro.gsfc.nasa.gov}}.

The simulation is initialized with the following inputs: a list of
observers and a target list of known planet-bearing stars.  Each
observer has an associated location (latitude and longitude) and
weather (average fraction of clear/cloudy nights per year).  Each
target has an associated position (RA and Dec), period $P$, estimated
transit probability $\Ptrans$ calculated from
equation~\ref{Ptrans-eq}, and a transit ephemeris.  The actual
ephemeris for any given system can be obtained from fits to existing
RV data, but for simplicity in this simulation, transit ephemerides
are generated randomly instead.  In cases where a single star hosts
multiple planets, it is listed in the target list with multiple
entries. Once observers and targets have been set up, several
record-keeping logs are also initialized.

The first Monte Carlo step is to assign which targets will host real
transits in the simulation.  The program calculates a true/false
condition for each target based on its $\Ptrans$.  The simulation then
enters its main loop which proceeds through an observing campaign
night-by-night; within each night there is nested a loop which
proceeds through the observer list one-by-one.  That inner loop over
observers proceeds as follows.  

Before assigning a target to an observer, the simulation first must
determine if the weather is favorable for the night.  While
season-based weather patterns have not been figured into the model,
the fraction of clear and cloudy nights at each observing location is  
known\footnote{For locations in the US, such data are available from
\url{http://www.ncdc.noaa.gov/oa/climate/online/ccd/cldy.html}}.
Using these probabilities, the night's weather for each location is
determined in Monte Carlo fashion. If clear or cloudy, all observers
common to the location will be affected identically. If the weather is
determined to be partly cloudy, each observer must be dealt with
independently, allowing for the possibility that some observers in a
given location will be able to observe while others are not.

For each observer, the Julian date (JD) of sunrise and sunset at the
observer's location are calculated for the current date in the
campaign.  The JD of sunset, the position of each target, and the
observer's location are used to calculate airmasses for each target in
the target list.  Any targets that pass the airmass cutoff (2.5 in our
simulations) at sunset are then checked to ensure they will pass the
airmass limit for at least four hours.  Thus a night of data will
consist of at least four hours of time-series photometry, and we allow
it to be as long as nine hours if the target is up (and the Sun is
down).

With the narrowed list of targets that are up, the simulation next
checks to see if any targets are near transit.  We assume that transit
ephemerides are accurate to within \mbox{$\sim$5\%} of an orbit.  If
the observer's night overlaps at all with this margin of error for a
target that is up, then that observer is assigned to that target.  The
likelihood of \textit{two} observable stars being at transit in a
given night is low, but if such a case does occur, the observer is
assigned to the higher transit probability.  Although there will be
fewer opportunities to observe the longer-period planets, we feel it
is justified to concentrate resources on targets where the
probability of successfully observing a transit is higher.

Thus, only if an observer has a target near-transit and favorable
weather, the simulation generates the night's photometry via Monte
Carlo.  Photometry is generated with an arbitrary zeropoint and
Gaussian noise.  Additionally, for those targets that the simulation
has randomly designated ``real'' transits, a simple linear
ingress/egress and transit depth based on \hdb\ are input into the
photometry.  We scale the ingress/egress times with the period of the
planet in order to simulate long-period transits.  The transit depth
and photometry noise amplitude are always set to match our template
data (plotted in Figure~\ref{transit-fig}) from the system described
in \S\ref{small-telescopes-sec}.  

An observer's photometry for a night is very simply analyzed by
calculating the Spearman's rank correlation for the data.  Basically,
this determines whether a linear trend has been detected between the
beginning and end of a night's data and returns a confidence-level for
such a trend \citep{nr1992}.  This simple analysis fits the
overwhelming majority of cases where portions of ingress or egress
have been observed, but will fail in the exceedingly rare case where
the transit is perfectly centered in the night's time window.

The process of the preceding five paragraphs is repeated for each
observer in the observer list, and this constitutes one night of the
campaign (the inner loop).  The simulation then increments its
internal ``calendar'' by one day and proceeds again --- this repeats
until the campaign ends (the outer loop).

For each target, the simulation keeps track of the number of
no-correlations, \nsig{2} correlations, and \nsig{4} correlations
observers have seen.  Any \nsig{2} correlations are used to give a
target a ``free pass'', allowing it to stay on the target list until
more definite observations can be made.  However, since the goal of
any campaign is to observe as much of the target list as possible,
targets must be eliminated from the list.  Thus, a limit is set on the
number of times a \nsig{4} correlation is shown before a star is
dropped.  This limit is generally low, for two reasons: in a real run,
someone with access to a professional observatory will begin follow-up
work, and additionally, in multiple realizations the \nsig{4}
correlation was associated with false positives in less than
1\% of all cases.  Similarly, a limit is set on the number of times a
target can produce a non-detection (no-correlation) before it is
dropped from the list.  This limit is generally fairly high, to avoid
dropping an actually-transiting target simply because the transit
occurred outside an observing window.  Both drop limits scale with
target planet period, as the 5\% accuracy of ephemerides leads
to exceedingly long observing windows for targets with periods as
short as 100 days, leading necessarily to a higher number of
non-detections.

In addition to this scorecard for each target, the simulation also
generates many other record-keeping files, such as a log of the
weather and observations of each observer for each night, and every
photometry file that is generated.  Certain variables of interest are
also tracked, such as the summed transit expectation value from the
remaining target list, updated whenever a target has been successfully
eliminated from further observing.

\section{Results}

We have modeled several different configurations of observers, and a
number of target lists. Addressed in this paper are five observer
scenarios: a lone, dedicated observer at Mt. Hamilton near San Jose,
CA; ten observers located at Mt. Hamilton; ten observers distributed
across the U.S.; twenty observers split between San Jose and Sydney,
Australia; and twenty observers distributed between eight worldwide
locations. For comparison, a second run of a single observer has been
made, but using a noise amplitude one-fifth that of the other runs, to
represent an astronomer with access to a professional observatory. In
each case, simulations are run for three target lists drawn from the
pool of planet-bearing stars. The selection is based on planetary
period, with maximum period cutoffs of 1000 days (essentially the
complete listing of extrasolar planets), 365 days, and 100 days. In
each case, 7 days is always the minimum period cutoff.

During trial runs, the \nsig{4} correlation and no-correlation limits
were adjusted to minimize the number of false positives/negatives,
while not making runs overly long. The values we used for these were
such that there were no false positives among the amateurs, and two
for the professional observer, for which the drop limits were lowered
to match the reduced noise factor. While several actual transits were
missed, these were due to incompleteness at the end of the run, not
due to observers incorrectly ruling them out.

For each pair of observer list and target list, the simulation has
been run 20 times to provide adequate statistics, and was analyzed to
find a number of quantities, most importantly list
completeness. Though it is possible to run the code until every target
has been adequately observed, this usually leads to prohibitively long
runs. Often, the end of a run is dominated by long period planets
which provide few opportunities for observing, and depending on
observer locations, may not \textit{ever} be observed. Rather than
running to absolute completion, we chose a fixed length for the
run. In each case presented here, runs start near the end of summer,
2003, and end in January 2008. This generally means that the observers
will not have completed the entire target list, but still provides a
good demonstration of the differences between observer configurations.

\begin{deluxetable}{rccc}
\tablecolumns{4}
\tablewidth{0pt}
\tablecaption{Simulation Results\label{table}}
\tablehead{
\colhead{} & \multicolumn{3}{c}{Period upper
  limit for target list} \\ \cline{2-4}
\colhead{Observer Configuration} & \colhead{100d} & \colhead{365d} &
\colhead{1000d} }
\startdata
1 observer, San Jose & 0.5 $\pm$ 1.5 & 0.4 $\pm$ 1.0 & 0.5 $\pm$ 0.7\\
1 professional observer, San Jose & 55 $\pm$ 7 & 45 $\pm$ 4 & 47 $\pm$ 4 \\
10 observers, San Jose & 62 $\pm$ 5 & 52 $\pm$ 3 & 41 $\pm$ 3 \\
10 observers, U.S.A. & 65 $\pm$ 6 & 56 $\pm$ 3 & 44 $\pm$ 3 \\
20 observers, San Jose \&  Sydney & 74 $\pm$ 6 & 64 $\pm$ 4 & 50 $\pm$ 4 \\
20 observers, worldwide & 95 $\pm$ 3 & 76 $\pm$ 4 & 56 $\pm$ 3 \\
\enddata
\tablecomments{Percent of target list completed by end of run for each
  observer configuration and target list.}
\end{deluxetable}

\begin{figure*}[h]
\epsscale{2.0}
\plotone{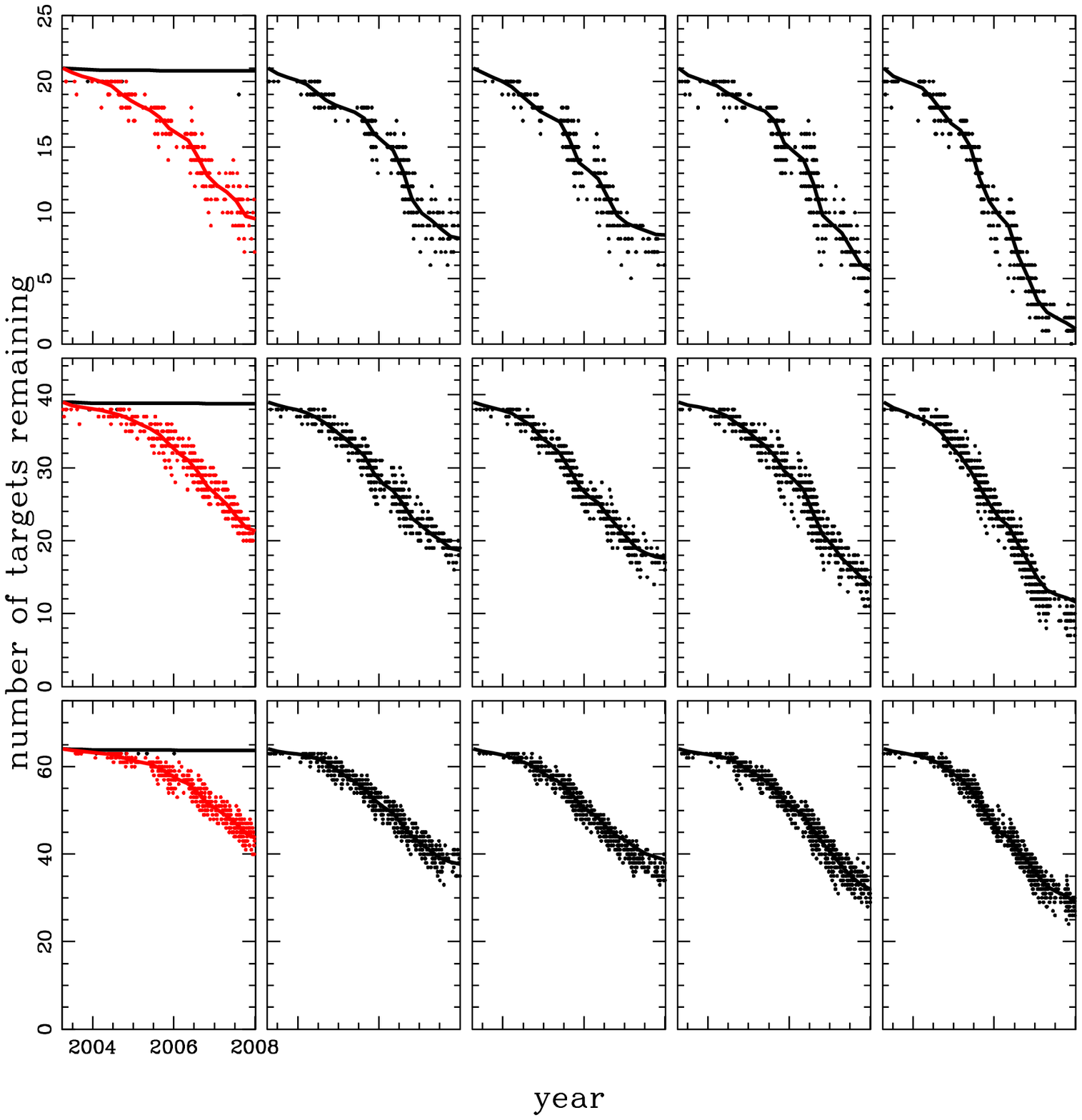}
\caption{
Simulated performance of sets of observers, showing number
of targets remaining vs.\ time.  The five columns of figures, from left
to right, show survey results from (1) A single observer at
Mt. Hamilton, CA, (2) ten observers on Mt. Hamilton, (3) ten observers
spread across the continental US, (4) twenty observers divided between
two locations, one in each hemisphere, and (5) twenty observers
distributed world-wide. The three rows, from top to bottom, show
results from period-limiting the target list at 100, 365, and 1000
days, respectively.  The red curve represents a single observer on
Mt. Hamilton, but with access to a telescope providing photometry
$\sim$5 times as accurate.  A point is generated for every date on
which a target was dropped from the list; points from all 20
realizations are shown.  The line is a smoothed average over
realizations.
\label{targets-remain-fig}
}
\end{figure*}

The final list completeness is tallied in
Table~\ref{table}. Additionally, plots of targets remaining to be
classified vs. time are shown in Figure~\ref{targets-remain-fig},
where we have recorded the JD on which targets were dropped and the
number of targets remaining afterward.  A smoothed average curve is
superimposed upon the target vs.\ time data from all 20 runs. Number of
observers increases from left to right, as does the average physical
separation between observers. The size of the target list increases
downwards. The red curve is the test case for our professional
astronomer.

In all cases, the curves show a characteristic delay time, in which
observers begin to classify targets, but have not had enough nights to
remove targets from consideration. At the \mbox{$\sim$2} year mark,
targets begin to become saturated, and the list of remaining targets
shortens rapidly. Eventually, most targets are removed, and the curve
flattens out as only the most difficult targets remain. As one would
expect, having more observers increases the number of targets it is
possible to cover by the end of the run, and additionally, it is
easily seen in the 100 day list and 365 day lists that observers more
evenly spread in longitude do a slightly better job completing the
target list. Also, note that while a lone amateur observer virtually
never amasses enough data to drop a target, even an observer with full
access to a professional-grade telescope can do no better than 10
observers clustered around the same location.

\begin{figure*}[h]
\epsscale{2.0}
\plotone{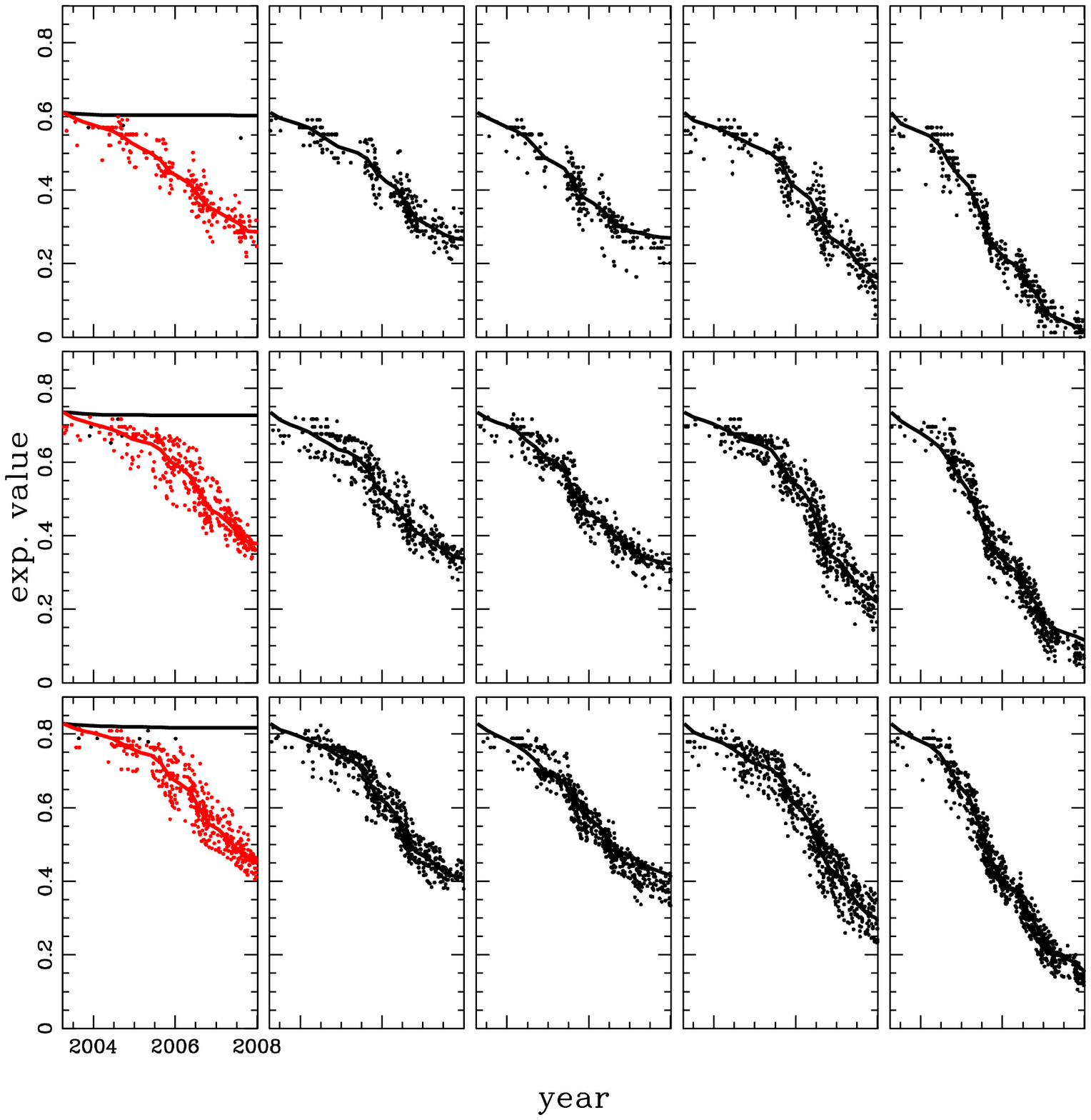}
\caption{
Simulated performance of sets of observers, showing
remaining transit expectation value vs.\ time. The five columns and
three rows of figures are as in Figure~\ref{targets-remain-fig}.
\label{exp-value-remain-fig}
}
\end{figure*}

The differences are better illustrated when we plot not the
number of targets remaining, but the summed transit expectation value
remaining on the target list (see
Figure~\ref{exp-value-remain-fig}). It is easier to see the
effect of longitudinal spread among observers. The second and fourth
columns represent observers concentrated in one and two locations,
respectively, and show almost three years before significant target
completeness begins to show. By contrast, the third and fifth columns
represent observers located over a larger spread in longitude, and
they begin to make progress fully a year before their
counterparts. Globally, the optimal configuration seems to be a large
number of observers with maximal spread in longitude/latitude. From
our data, we show that this configuration of observers can cover
roughly 40\% more of the probability-weighted target list than even
the professional observer.

\section{Conclusion}

These Monte Carlo simulations are a feasibility study that
demonstrates the efficacy of the \url{transitsearch.org} project.
They differ from reality in important ways.  For instance, in the
simulations a simple rank-correlation analysis is applied to
individual observers' data, for efficiency.  In reality, multiple
observers' data will overlap, and arbitrarily sophisticated techniques
(along with eyeballs) will be brought to bear on anything that appears
interesting.  A shortcoming of the simulations is that the photometric
noise and transit signal are based on the data from our demonstration
observatory described in \S\ref{small-telescopes-sec}.
Intermediate-period planets will likely have smaller radii than \hdb,
and hence the transit signal will not be as deep.  However, many
\url{transitsearch.org} ``amateurs'' obtain photometry comparable to
data from our setup --- and the simple improvements mentioned in
\S\ref{small-telescopes-sec} and \citet{cast2000} will increase
photometric precision.  In addition, the longer timescale of
intermediate-period transits will allow for more binning of the
photometry, further increasing precision.  Lastly, the transit depth
is also very sensitive to the stellar radius, which varies
significantly in the target list, but has not been considered here.
Nevertheless we feel that this feasibility study is at least a
reasonable demonstration of our strategy.

These simulations show that while a single observer campaign is
capable of discovering transits, this observer will generally leave
\mbox{30-50\%} of the sky uncovered. Not only can multiple observers
better cover the sky, they can also cover it more
quickly. Additionally, the data reveal the importance of having not
only multiple observers in multiple locations, but also in ensuring
that the observers cover a wide range of longitudes in both
hemispheres. Note, for instance, the difference in time to completion
between the case where 10 observers are located in both San Jose and
Sydney, and the case where 20 observers are scattered across nine
worldwide locations. It is apparent that longitudinal coverage is
important. One naturally expects that weather will be a key factor in
determining time to completion, as it will most dramatically affect
the length of a run that is confined to a single location. But the
spread in longitude proves equally important, as one might guess from
the process of viewing eclipses on Earth. Both timing \textit{and}
location are everything.

Most current work on transits is divided into two categories, our
Mount Hamilton case (the single dedicated observer), and studies like
OGLE which rely on time sequenced, wide-field snapshots that detect
possible transits. However, we have shown that a single observer is at
a disadvantage, no matter how powerful the telescope, while wide-field
surveys suffer from false positives associated with binary stars, and
additionally provide poor targets for follow-up radial velocity
work. In the end, even confining a search to the known extrasolar
planets produces a long list of potential targets that proves
difficult to work through.  We have shown that by handing the bulk of
observing work to a dedicated team of observers with good longitudinal
coverage, we may ensure that when a transit is expected to occur,
there is \textit{always} someone watching, and that this team will
prove competitive with any other transit search venture.

\end{document}